\documentclass[onecollarge,natbib]{svjour2}
\bibpunct{[}{]}{;}{n}{}{,} 
\smartqed  
\usepackage{graphicx}
\usepackage{amsmath, amssymb}
\usepackage{color}
\journalname{Few-Body Systems}

\begin{document}

\title{Impenetrability in Floquet scattering in one dimension}

\author{A.~G.~Volosniev \and D.~H.~Smith}


\institute{A.~G.~Volosniev  \at
              Institut f{\"u}r Kernphysik, Technische Universit{\"a}t Darmstadt, 64289 Darmstadt, Germany   
  \newline
 \email{volosniev@theorie.ikp.physik.tu-darmstadt.de}    
           \and
           D.~H.~Smith \at Department of Physics, The Ohio State University, Columbus, OH 43210, USA  
}


\date{Received: date / Accepted: date}

\maketitle

\begin{abstract}
We study the scattering off a time-periodic zero-range potential in one spatial dimension. We focus on the parameter regions that lead to zero-transmission probability (ZTP). For static potentials, ZTP leads to fermionization of distinguishable equal-mass particles. For time-periodic potentials, fermionization is prevented by the formation of evanescent waves. 
\end{abstract}

\section{Introduction}

We recently witnessed the exciting development of experimental techniques that allow physicists to adjust the scattering properties of cold atoms at will, in particular, to tune the scattering length, $a$, using Feshbach resonances~\cite{chin2010}. The ability to control $a$ enabled the experimental realizations of paradigmatic theoretical predictions in condensed matter physics. A celebrated example in one spatial dimension is the Bose-Fermi mapping, which was proposed in 1960~\cite{girardeau1960} and realized in 2004~\cite{tg2004,Kinoshita1125}. It implies that the probability density function of a trapped gas of impenetrable bosons is given by the probability density function of a gas of spin-polarized (or spinless) fermions, which allows one to understand various aspects of strongly-interacting trapped systems~\cite{girar2001, deuret2008,volosniev2014}. To observe this correspondence one works with impenetrable particles (i.e., particles whose scattering assumes ZTP), which are produced by tuning the effective one-dimensional (1D) interaction strength, $g_0$, related to $a$~\cite{olshanii1998}. For static potentials, impenetrability occurs only if the energy scale given by $g_0$ is much larger than any other energy scale of the problem, in other words if $g_0\to\infty$.  In this contribution we discuss another route towards ZTP that requires precisely tuned time-dependent zero-range potentials~(cf.~\cite{roger1992}). Our study is motivated by the possibility to explore this route with cold atoms in the modulated external magnetic field (cf.~\cite{wieman2005}).

Two- and few-body physics with time-periodic short-range  potentials has been widely discussed theoretically in the context of particle tunneling \cite{landauer1982, landauer1985}, electron transmission \cite{roger1992,wagner1995,maschke1998,li1999}, quantum irregular models \cite{richter01}, molecule association~\cite{burnett2007, braaten2015}, engineering of strongly-interacting samples~\cite{hudson2015, hudson2016,dane_thesis}, Efimov physics~\cite{petrov2017},  etc. A standard theoretical approach for studying quantum phenomena in the presence of time-periodic potentials is the Floquet formalism (for a review see, e.g., Ref. \cite{hanngi1998}), which we also use in our study.

The paper is organized as follows: In Sec.~\ref{sec:zero} we study the scattering from a zero-range potential and the formation of metastable states. In Sec.~\ref{sec:concl} we make some concluding remarks. In the appendix we provide a theoretical framework for a scattering off a time-periodic potential.

\section{Scattering off a time-periodic zero-range potential in one dimension}
\label{sec:zero}

{\it \bf Formulation.} We consider low-energy atom-atom scattering in 1D, assuming a time-periodic interaction potential, which can be engineered by modulating an external magnetic field, see, e.g.,~\cite{wieman2005}. We model the zero-range atom-atom interaction with a Dirac-delta function potential~\cite{olshanii1998}.  The center-of-mass momentum is conserved, and the Schr{\"o}dinger equation that describes relative motion reads
\begin{equation}
i\frac{\partial}{\partial t}\Psi(x,t)=\left(-\frac{\partial^2}{\partial x^2}+g_0\delta(x)+g_1\cos(\omega t)\delta(x)\right)\Psi(x,t).
\label{eq:schr0}
\end{equation}
We have chosen the units such that the reduced mass equals to one half, and $\hbar=1$; without loss of generality we assume that $g_1>0$. We solve this equation using plane waves as a basis. To specify that the incomining flux is $e^{ipx}$, Eq.~(\ref{eq:schr0}) should be supplemented by the boundary conditions at $x\to\pm\infty$, which define the wave function everywhere except at $x=0$, because the interaction potential is of zero range. The wave functions at $x \neq 0$ can therefore be written as
\begin{equation}
\Psi_{x>0}=\sum_{n=-\infty}^{\infty}e^{-ip_n^2 t}\left(\delta_{n,0}+B_n\right)e^{ip_n x},\quad \Psi_{x<0}=\sum_{n=-\infty}^{\infty}e^{-ip_n^2 t}\left(\delta_{n,0}e^{ip_n x}+B_n e^{-ip_n x}\right),
\label{eq:bound_cond_0}
\end{equation}
where $p_n\equiv\sqrt{p^2+n\omega}$, $p$ is the incoming momentum and $\delta_{i,j} $ is the Kronecker delta. The function $\Psi_{x\neq 0}$ contains the incoming flux $e^{ip x}$ and all of the allowed outgoing (evanescent) plane waves, i.e., the waves that have the energies $p^2+n\omega$ [$n=0,\pm 1,\pm 2, \dots$] -- note that the energy is not conserved because the interaction depends on time. See Appendix~\ref{sec:formalism} for more detail. 

\vspace*{1em}

\noindent {\it \bf Formal solution.} To find the coefficients $B_n$, we match the $x<0$ and $x>0$ solutions using the boundary conditions at $x=0$  given by the delta-function interaction potential. This leads to the recurrence relation (see also~\cite{roger1992} and Appendix~\ref{sec:formalism})
\begin{equation}
2 p_n C_n=2 p\delta_{n,0}-ig_0 C_n-\frac{i g_1}{2}\left(C_{n+1}+C_{n-1}\right),
\label{eq:set_zero}
\end{equation}
where $C_n=B_n+\delta_{n,0}$. This relation reflects the coupling between the different energy components of the wave function induced by the time-periodic potential. The coefficients $C_{n}$ with $|n|\gg 1$ vanish (e.g., $C_{n+1}\simeq C_n/\sqrt{n}$ for $n\gg1$) to ensure that the outgoing and the incoming fluxes are equal (cf. Appendix \ref{sec:appA}). Therefore, we obtain a systematically improvable approximation by truncating the infinite sum in Eq. (\ref{eq:set_zero}) to a finite number of terms by setting $C_{|n|>n_{max}}=0$. The accuracy and validity of this truncation are examined  {\it a posteriori}, when we inspect the dependence of the results on $n_{max}$. Truncated Eq.~(\ref{eq:set_zero}) is simply a linear system. To solve it, we define the following matrix ($g_1\neq 0$): $M_{kl}=\frac{2 i}{g_1}(2p_{l-1-n_{max}}+ig_0)\delta_{k,l}-\delta_{k,l-1}-\delta_{k,l+1}$ for $k,l=1,2,...,d$, where the dimension of this matrix, $d=2n_{max}+1$, equals the number of energy components in the approximate solution for the wave function. The matrix $M$ is tridiagonal, which allows us to apply fast numerical routines to solve the problem. Moreover, some of the results can be obtained in a closed form. For example, the determinant and inverse (if $\mathrm{det}M\neq0$) of $M$ can be computed, see, e.g., Refs.~\cite{Mallik2001,ElMikkawy2004}. 
The inverse matrix $M^{-1}$ can be used to find a formal solution to Eq. (\ref{eq:set_zero}): $C_n=\frac{4p i}{g_1}(M^{-1})_{n+1+n_{max},1+n_{max}}$. Even though the inverse matrix can be written in a closed form, the final expression is cumbersome and does not give additional insight into the problem. Therefore, we refrain from showing it here. Instead we discuss regions with ZTP.

\vspace*{1em}

\begin{figure}
\centering
\includegraphics[scale=0.7]{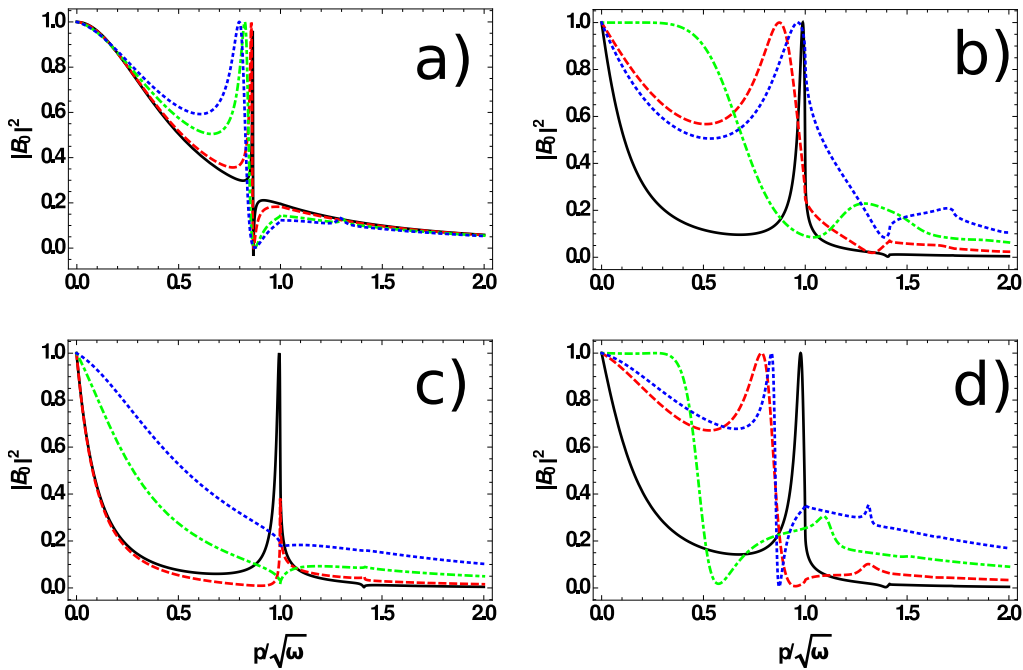}
\caption{Panel {\textbf a)}: The reflection probability $|B_0|^2$ as a function of $p/\sqrt{\omega}$ for $g_1/\sqrt{\omega}=0.2$ -- black solid curve, $g_1/\sqrt{\omega}=0.4$ -- red dashed curve,  $g_1/\sqrt{\omega}=0.8$ -- green dot-dashed curve, and $g_1/\sqrt{\omega}=1.0$ -- blue dotted curve. For all curves $g_0/\sqrt{\omega}=-1$. Panel {\textbf b)}: $|B_0|^2$ for $g_1/\sqrt{\omega}=1.5$ -- black solid curve, $g_1/\sqrt{\omega}=2.5$ -- red dashed curve, $g_1/\sqrt{\omega}=3.5$ -- green dot-dashed curve, and $g_1/\sqrt{\omega}=4.5$ -- blue dotted curve. For all curves $g_0=0$. Panel {\textbf c)}: $|B_0|^2$ for $g_0/\sqrt{\omega}=0.1$ -- black solid curve, $g_0/\sqrt{\omega}=0.5$ -- red dashed curve,  $g_0/\sqrt{\omega}=1$ -- green dot-dashed curve, and $g_0/\sqrt{\omega}=1.5$ -- blue dotted curve. For all curves $g_1/\sqrt{\omega}=1.5$. Panel {\textbf d)}:  $|B_0|^2$ for $g_0/\sqrt{\omega}=-0.1$ -- black solid curve, $g_0/\sqrt{\omega}=-0.8$ -- red dashed curve,  $g_0/\sqrt{\omega}=-1.4$ -- green dot-dashed curve, and $g_0/\sqrt{\omega}=-2$ -- blue dotted curve. For all curves $g_1/\sqrt{\omega}=1.5$.}
\label{fig:fig1new}
\end{figure}

\noindent {\bf Zero transmission.} For simplicity, for the analytic analysis we assume that $p^2\in (0,\omega)$. In this case zero transmission requires vanishing $C_n$ for $n\geq0$. The recurrence relation in Eq.~(\ref{eq:set_zero}) reads
\begin{equation} 
C_{-1}=4p/(i g_1),\quad C_{-2}=2i(2p_{-1}+i g_0)C_{-1}/g_1, \quad \dots \; .
\end{equation} 
As we have seen, the coefficients $C_n$ decrease very rapidly with increasing number $|n|$, so as a first approximation for a weak coupling $g_1$ we can put $C_{-2}=0$, which leads to $p_{-1}\simeq-ig_0/2$. Below we show that this approximation is indeed very accurate if $g_1\to0$. The full reflection can be achieved if $g_0<0$ and $\omega=p^2-E_b$, where $E_b=-g_0^2/4$ is the ground state energy for $g_1=0$. This is illustrated in Fig.~\ref{fig:fig1new}{\textbf a)}, which resembles a picture of a traditional Fano-Feshbach resonance. In our case the coupling to a long-lived bound state is achieved by means of a time-periodic potential, which reveals the energy of the ground state of the underlying static potential (cf.~\cite{wieman2005, maschke1999}). 

To estimate the accuracy of our approximation (i.e., $C_n = 0$ for $n\leq -2$), we assume that $C_{-2}\neq0$ and $C_{-3}, C_{-4},...=0$. This leads to the condition 
\begin{equation}
2 p_{-1}\simeq-i g_0-\frac{g_1^2}{4 (2 p_{-2}+ig_0) } \to \frac{p^2}{\omega}\simeq 1-\frac{g_0^2}{4\omega}-\frac{|g_0| g_1^2}{8\omega^{3/2}\left(\sqrt{\frac{g_0^2}{\omega}+4}+\frac{g_0}{\sqrt{\omega}}\right)}.
\label{eq:correction_1}
\end{equation}
Equation~(\ref{eq:correction_1}) shows that if $g_1$ is much smaller than $g_0$, then we reproduce the result above, hence, $C_{-2}$ can be indeed put to zero. Note that the inclusion of $g_1$ shifts the position of the resonance towards smaller values of $p/\sqrt{\omega}$ (see Fig.~\ref{fig:fig1new}{\textbf a)}).
The correction~(\ref{eq:correction_1}) also reveals that zero transmission can be reached even without an underlying bound state, i.e., for $g_0=0$. In this case the condition reads $16 p_{-1}p_{-2}  \simeq -g_1^2$ (e.g., for $g_1 \simeq \omega$ ZTP occurs at $\omega\simeq p^2$; see Fig.~\ref{fig:fig1new}{\textbf b)}). 

To summarize the discussion in this subsection: ZTP can be achieved with and without an underlying bound state. The coupling between a well-defined bound state (i.e., the limit $g_1\to 0$) and a continuum, realizes the Fano scenario for the appearance of resonance features~\cite{fano1961}. This resonance has an asymmetric shape. When, $g_0$ becomes comparable or smaller than $g_1$, which can be seen as strong coupling, the discrete level becomes embedded into the continuum producing the Breit-Wigner shape of the resonance. This is what we see for $g_0=0$; see Fig.~\ref{fig:fig1new}{\textbf b)}.

\vspace*{1em}

\noindent {\bf Numerical solution}.  We also solve Eq.~(\ref{eq:set_zero}) (exactly) numerically and plot the reflection coefficient, $|B_0|^2$, in Fig.~\ref{fig:fig1new}. The observable $|B_0|^2$ determines the probability for the incoming wave $|p\rangle$ to be reflected in the same mode. Therefore, $|B_0|^2=1$ corresponds to zero transmission. In Fig. \ref{fig:fig1new} we demonstrate $|B_0|^2$ as a function of $p/\sqrt{\omega}$ for different values of $g_1$ and $g_0$. The resonance features in panels {\textbf a)} and {\textbf b)} are discussed above. We also point the reader to the plateau for $g_1/\sqrt{\omega}=3.5$ and $g_0=0$; see Fig.~\ref{fig:fig1new}{\textbf b)}.  This plateau ensures that the wave packet is fully reflected if the incident momentum does not exceed $p/\sqrt{\omega}\simeq 0.5$. Panel {\textbf c)} displays the effect of $g_0>0$.  As expected, very small $g_0$ does not noticeably affect the scattering properties. However, for larger $g_0$ the shape of the resonance feature changes and becomes washed out.  Panel {\textbf d)} shows the ``transition" from the Breit-Wigner to the Fano shape of the resonance driven by $g_0<0$. Note that there is a zero-transmission plateau in the transition region, similar to the one in Fig.~\ref{fig:fig1new}{\textbf b)}. However, unlike the scattering off a static potential, this plateau does not lead to fermionization in trapped systems of bosons [that scatter with $p/\sqrt{\omega}  \lesssim 0.4$]. Indeed, besides the reflected wave there will be evanescent waves (see Eq.~(\ref{eq:bound_cond_0})) that lead to the formation of metastable states (see the next section). These states are not populated in the scattering of two spinless fermions and thus the Bose-Fermi mapping cannot be used to describe them.

\vspace*{1em}

\noindent {\bf Formation of metastable states.} The appearance of the resonant features in Fig.~\ref{fig:fig1new}{\textbf a)} is connected to the existence of a bound state for $g_1=0$. If $g_1>0$ there is a coupling between an incoming wave and this discrete level. This coupling leads to the observed resonance feature. During the resonance collision particles populate the `bound' state, which then decays due to the coupling to the continuum. We refer to this resonant population of the bound state as the formation of a metastable state. To investigate this process, we assume that the initial state is not a plane wave but the wave packet $\Psi(x,0)$ well-localized within the first Floquet band, i.e., $0<p^2<\omega$. 
The time-dependent wave function $\Psi(x,t)$ is then easily written using results for plane waves
\begin{equation}
   \Psi(x,t)=\Psi_0(x,t)+ \left\{
                \begin{array}{ll}
                  \sum_n \int \mathrm{d}p\phi(p) B_n e^{-ip_nx-ip_n^2t}, x<0\\
                  \sum_n \int \mathrm{d}p\phi(p) B_n e^{ip_nx-ip_n^2t}, x>0
                \end{array}
              \right.
\end{equation}
where $\Psi_0(x,t)$ is the non-scattered wave. We assume that the initial state and $\phi$ have the form
\begin{equation}
\Psi(x,0)=\left(\frac{\Delta}{2\pi}\right)^{1/4} e^{ip_0 x} e^{-\frac{\Delta x^2}{4}} \; \to \; \phi(p)=\left(\frac{1}{2\Delta\pi^3}\right)^{1/4}e^{-\frac{(p-p_0)^2}{\Delta}},
\label{eq:wave_packet}
\end{equation} 
here the normalization condition is $\langle \Psi|\Psi \rangle=1$.  For convenience, we use $\phi(p)=0$ for $p<0$ and $p^2>\omega$ (this prescription does not affect the results for well-peaked wave packets ($p_0/\Delta\gg 1$) considered below).

We calculate the overlap of the wave function with the bound state present in the static potential
\begin{equation}
F(t)=\sqrt{\frac{|g_0|}{2}}\int \mathrm{d}x e^{-\frac{|g_0 x|}{2}}\Psi(x,t).
\end{equation}
The function $|F(t)|^2$ represents the probability to observe a bound state if at the time $t$ the coupling $g_1$ is switched off. The function $F(t)$ reads
\begin{equation}
F(t)=\sqrt{2 |g_0|}\int \mathrm{d}p\phi(p)\left(-\frac{ipe^{-i p^2 t}}{\frac{g_0^2}{4}+p^2}+\sum_n \frac{C_ne^{-ip_n^2 t}}{\frac{|g_0|}{2}-ip_n}\right).
\end{equation} 
\begin{figure}
\centering
\includegraphics[scale=0.57]{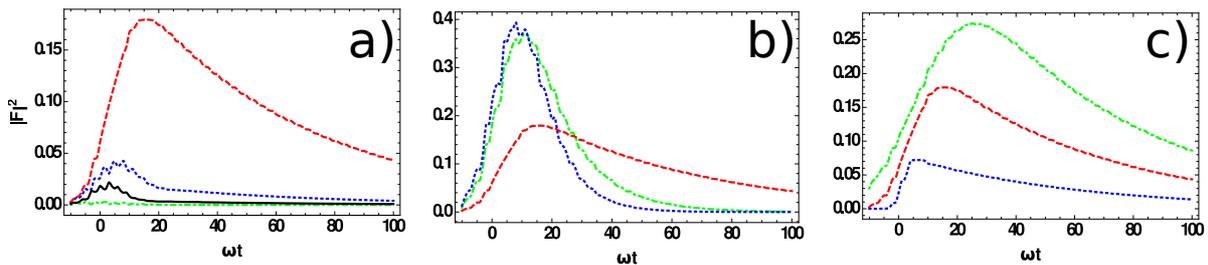}
\caption{ Panel {\textbf a)}: The overlap  $|F|^2$ as a function of $\omega t$ for $p_0=\sqrt{\omega}/2$ --  green dot-dashed curve, $p_0=3 \sqrt{\omega}/4$ -- blue dotted curve,  $p_0= \sqrt{3\omega/4}$ (resonance) -- red dashed curve, and $p_0=\sqrt{\omega}$ -- black solid curve.  The other parameters are: $g_0/\sqrt{\omega}=-1$, $g_1/\sqrt{\omega}=0.4$ and $\Delta=\omega/100$, see the text for details. Panel {\textbf b)}: $|F|^2$ for $g_1/\sqrt{\omega}= 2/5$  -- red dashed curve, $g_1/\sqrt{\omega}=4/5$ --  green dot-dashed curve, and $g_1= \sqrt{\omega}$ -- blue dotted curve.  The other parameters are: $g_0/\sqrt{\omega}=-1$, $p_0$ is given by the resonance condition (see the text) and $\Delta=\omega/100$. Panel {\textbf c)}:  $|F|^2$  for   $\Delta=\omega/100$ -- red dashed curve, $\Delta=\omega/400$ --  green dot-dashed curve, and  $\Delta=\omega/10$ -- blue dotted curve. The other parameters are: $g_1/\sqrt{\omega}$= 2/5, $g_0/\sqrt{\omega}=-1$, and $p_0$ is fixed by the resonance condition (see the text).}
\label{fig:fig2new}
\end{figure}
Let us start our investigation of $F$ with the case $g_1/\sqrt{\omega}=0.4$ and $g_0/\sqrt{\omega}=-1$; see the (red) dashed line in Fig. \ref{fig:fig1new}{\textbf a)}. These parameters lead to a narrow resonance at $p^2\simeq \omega-\frac{g_0^2}{4}$ (see above) whose width can be determined from $\mathrm{det}M=0$; cf.~\cite{roger1992, li1999}. For small $g_1$ we find the first pole at approximately $p^2\simeq \omega-\frac{g_0^2}{4}-i \frac{g_1^2|g_0|}{16\sqrt{\omega}}$, where only the leading order terms in both the real and imaginary parts are kept. To investigate the scattering close to the resonance, we 
use a narrow wave packet with  $\Delta=\omega/100$, which is of the order of the width of the resonance. We plot our findings in Fig.~\ref{fig:fig2new}{\textbf a)}; the plot shows that if $p_0$ is far from the resonance, the metastable state is not populated and $F(t)\simeq 0$. For the resonant value of $p_0$, there is a significant probability to populate a `bound state', which will then decay. One can easily show that the decay is exponential, i.e., $\sim e^{-\Gamma t}$; moreover, it is the same for all curves in Fig.~\ref{fig:fig2new}{\textbf a)}, and, thus, the decay can indeed be associated with a metastable state. The best fit to the exponential decay in the presented region  yields $\Gamma\simeq 0.11 g_1^2 |g_0|/\sqrt{\omega}$.
This result compares well with the prediction from $\det M =0$ discussed above, i.e., $\Gamma= g_1^2|g_0|/(8\sqrt{\omega})$. We continue our investigation (see Fig.~\ref{fig:fig2new}{\textbf b)}) by considering different values of $g_1$ with a wave packet peaked at the resonant position determined above. As expected, a stronger coupling $g_1$ leads to a larger overlap $F(t)$ and to a faster decay. Finally, we investigate the role of $\Delta$. To this end, we fix $p_0$ to the resonance position and vary $\Delta$; see Fig.~\ref{fig:fig2new}{\textbf c)}, which shows that narrow wave packets lead, as expected, to a larger overlap $F$.

\section{Conclusions}
\label{sec:concl}

We have considered the scattering off a time-periodic potential in one spatial dimension. This scattering has regions with zero-transmission probability even for finite values of the interaction parameters. This, however, does not imply fermionization of particles because of the evanescent waves, which lead to the appearance of metastable states. Unusual scattering properties brought by time-periodic potentials make it interesting to speculate on possible applications of the presented results. The sharp decrease of the green dot-dashed curve in Fig.~\ref{fig:fig1new}{\bf d}) suggests  that by periodically modulating the shape of the trap, one might engineer one-body physics with resonant transport properties. A time-periodic change of trap boundaries might even cool the system due to the preferential escape of high-energy atoms. In general, the energy of the Hamiltonian with one- or two-body time-periodic potentials is not conserved, thus, these potentials can be used to change the total energy of the system. We checked that there are parameter regions for which the total energy is decreased after the scattering event, which also might lead to cooling of many-body systems. However, a thorough study is required to check this statement. Another interesting future investigation will be dedicated to the long-lived states discussed in the previous section. These states might enhance the role of three-body effects, and it will be interesting to investigate this few-body physics in cold gases. Such an investigation might also require the inclusion of a harmonic trap, as the confinement introduces another time scale, which might allow for the appearence of other phenomena, such as the parametric resonance~\cite{volosniev2015pol,ebert2016,smith2016}.

{\small {\bf Acknowledgments} A.~G.~V. gratefully acknowledges the support of the Humboldt Foundation.}

\appendix

\section{Appendix}
\label{sec:formalism}

\subsection{Floquet Formalism}

{\bf Scattering in time-dependent picture}. Here, for convenience of the reader, we discuss the time-dependent Schr{\"o}dinger equation with a general short-range time-periodic potential
\begin{equation}
i\frac{\partial}{\partial t}\Psi(x,t)=H(t)\Psi(x,t),\qquad H(t)=H_0+W(x,t),
\label{eq:schr}
\end{equation}
where the operators in the coordinate representation are $H_0=-\frac{\partial^2}{\partial x^2}$, and $W(x,t)=\sum_n  e^{-\frac{2 i \pi n t}{T}} W_n(x)$. We assume that the incoming square-integrable wave packet is $\Psi_0(x,t)$, i.e., $\Psi(x,t\to-\infty)=\Psi_0(x,t)$. Obviously, if $g_n=0, \forall n$ then the time propagation is $\Psi_0(x,t)=e^{-i H_0 t}\Psi_0$, where $\Psi_0\equiv\Psi_0(x,0)$. Therefore, the effect of scattering is deduced by comparing $\Psi(x,t)$ with $\Psi_0(x,t)$. 

To proceed we note that there is a unitary operator $U(t,s)$ that determines the time evolution
\begin{equation}
\Psi(t)=U(t,s)\Psi(s).
\end{equation}
The formal properties of $U(t,s)$ are discussed in Ref. \cite{yajima1977}.
From now on we reserve the letters $t$ and $s$ for time, also when it does not cause confusion we omit the coordinate variables. For convenience, we set $t=0$ to be a reference time and introduce $\Psi\equiv\Psi(x,0)$ such that $\Psi(t)=U(t)\Psi$, where $U(t)\equiv U(t,0)$. To compare $\Psi(t)$ and $\Psi_0(t)$ one can introduce the wave operator $\Omega(s)$, such that $\Psi=\Omega \Psi_0$, $\Omega=\Omega(0)$. This operator is defined as the limit
\begin{equation}
\Omega(s) \equiv \lim_{t\to-\infty}U^{-1}(t,s)e^{-i(t-s) H_0}.
\end{equation}
In scattering theory for time-independent potentials $\Omega$ is often called the M{\o}ller operator \cite{taylorbook}. This operator exists also for time-periodic short-range potentials as discussed in Refs.~\cite{yajima1977, howland1979}. One important property of $\Omega$ is called the intertwining relation:
\begin{equation}
U(t)\Omega=\Omega(t)e^{-i H_0 t}.
\end{equation}
We use this relation upon decomposing $\Psi_0$ in the eigenbasis of $H_0$. For convenience we refrain from using coordinate representation and use Dirac's notation for vectors instead: $|\Psi_0\rangle=\int_{-\infty}^{\infty} \mathrm{d}p \phi(p) |p\rangle$. Now if we apply the intertwining relation to this decomposition we obtain
\begin{equation}
\int \mathrm{d}p \phi(p)\left(U(t)\Omega |p\rangle -e^{-i p^2 t}\Omega(t)|p\rangle \right)=0.
\end{equation}
Since it should be valid for all possible initial wave packets described by $\phi(p)$ we  conclude that the integrand should be zero. By differentiating both sides with respect to time, we see that the only way to satisfy this condition is to assume that $|f_p\rangle(t)\equiv \Omega(t)|p\rangle$ obeys
\begin{equation}
\left(H(t)-i\frac{\partial}{\partial t}\right) |f_p\rangle(t) =p^2 |f_p\rangle(t).
\label{eq:floquet}
\end{equation}

Let us take a closer look at Eq. (\ref{eq:floquet}). According to the Floquet theorem $U(t+T,s+T)=U(t,s)$; see also Ref. \cite{yajima1977}. Therefore, $\Omega(T)=\Omega$, hence $|f_p\rangle(t+T)=|f_p\rangle(t)$ and $|f_p\rangle(t)=\sum_n e^{-\frac{2 i \pi n  t}{T}} |\tilde f_{pn}\rangle$. By inserting this ansatz function into Eq. (\ref{eq:floquet}) and projecting onto a particular mode we obtain 
\begin{equation}
(p^2+n \omega -H_0)| \tilde  f_{pn}\rangle= \sum_m W_{n-m}|\tilde f_{pm}\rangle,
\label{eq:set}
\end{equation}
where $\omega=2\pi/T$. This is a (infinite-dimensional) matrix equation, and therefore, for each $p^2$ there is an infinite number of solutions, which can be formally written as 
\begin{equation}
|\tilde f_{pn}\rangle = |p_n\rangle \alpha_n + \frac{1}{p^2+n\omega-H_0+ i\epsilon}\sum_{m}W_{n-m}|\tilde f_{pm}\rangle,
\label{eq:tildefpn}
\end{equation}
where $\alpha_n$ and $\epsilon\in\mathbb{R}$ are coefficients, and $p_n^2=p^2+n\omega$. We show below that the solution that satisfies the initial condition $\Psi(x,t\to-\infty)\to \Psi_0(x,t)$ has $\alpha_n=\delta_{n,0}$ and $\epsilon>0$. To this end
we write the formal solution to Eq.~(\ref{eq:schr}) 
\begin{equation}
|\Psi\rangle(t)=e^{-i H_0 t}\int_{-\infty}^{\infty}\mathrm{d}p\phi(p)\left(1-i \int_{-\infty}^t \mathrm{d}t' e^{i H_0 t'}W(t')\Omega(t')e^{-i H_0 t'}\right)|p\rangle.
\label{eq:psi_t}
\end{equation}
This equation yields $|\Psi\rangle$
\begin{equation}
|\Psi\rangle=\int_{-\infty}^{\infty}\mathrm{d}p \phi(p)\left(|p\rangle+\sum_{m,n}\frac{1}{p^2+(n+m)\omega-H_0+i\delta}W_{n}|\tilde f_{pm}\rangle\right).
\label{eq:psi_help1}
\end{equation} 
where $\delta$ is a small positive quantity. 	
To obtain $|\Psi\rangle$ we used the prescription $W(t)\to e^{\delta t}W(t)$, which is justified by noticing that for $t\to-\infty$ the square-integrable wave packet cannot be affected by the finite-range potential. Now if we look at Eqs.~(\ref{eq:tildefpn}) and~(\ref{eq:psi_help1}) and notice that
$|\Psi\rangle=\Omega |\Psi_0\rangle=\int \mathrm{d}p \phi(p)\sum_n |\tilde f_{pn}\rangle$ we deduce that $\alpha_n=\delta_{n,0}$ and $\epsilon>0$. At $t$ we have $|\Psi\rangle(t)=\int\mathrm{d}p \phi(p)e^{-ip^2t} \sum_{n}e^{-i\omega n t}|\tilde f_{pn}\rangle$. This provides us with the wave function $\Psi(x,t)$ for $x\to\infty$, which determines characteristics of transmission 
\begin{equation}
\Psi(x,t)=\Psi_0(x,t) -\frac{i}{2}\sum_{m,n}\int\mathrm{d}p\phi(p) \frac{e^{ip_nx - i p_n^2 t }}{p_n} \int\mathrm{d}x' e^{-ip_nx'} W_{n-m}(x')\langle x'|\tilde f_{pm}\rangle,
\label{eq:asymptotics}
\end{equation}
a similar expression can be derived for $x\to-\infty$.
Here we use the coordinate representation of the Green's function
\begin{align}
G(x,x';k^2)\equiv\langle x| \frac{1}{H_0-k^2-i\epsilon}|x'\rangle  = \frac{i}{2k}e^{ik|x-x'|}.
\end{align}

{\bf Scattering in time-independent picture}. In this subsection we consider Eq.~(\ref{eq:tildefpn}) that determines the properties of the scattering in time-independent picture in more detail. In the coordinate representation it reads
\begin{align}
\tilde f_{pn}(x)=\delta_{n,0}e^{i p_n x}- \sum_{m=-\infty}^{\infty} \int\mathrm{d}x' 
 G(x,x';p_n^2)W_{n-m}(x')\tilde f_{pm}(x'),
\label{eq:fpn}
\end{align} 
where $\delta_{m,n}$ is Kronecker's delta. This function at $x\to\infty$ has the form $\delta_{n,0}e^{ip_nx}+B_ne^{ip_nx}$ where 
\begin{align}
B_n=-\frac{i}{2 p_n}\sum_{m=-\infty}^\infty \int\mathrm{d}x' e^{-i p_n x'}W_{n-m}(x')\tilde f_{pm}(x')
\label{eq:an}
\end{align}
whereas at $x\to-\infty$ it has the form $\delta_{n,0}e^{ip_nx}+\tilde B_n e^{-i p_n x}$ with
\begin{align}
\tilde B_n=-\frac{i}{2p_n}\sum_{m=-\infty}^\infty \int\mathrm{d}x' e^{i p_n x'}W_{n-m}(x')\tilde f_{pm}(x'), 
\label{eq:bn}
\end{align}
Apparently, Eqs. (\ref{eq:fpn}), (\ref{eq:an}) and (\ref{eq:bn}) contain all information about the scattering process and can be used to derive Eq.~(\ref{eq:set_zero}) of the main text. It is worthwhile to notice that for a plane wave with a given $p^2$ the total probability to find a particle with the energies $p^2_n, n=0,\pm1,...$ is conserved in the scattering process (see the next subsection). This can be seen as the conservation of the quasi-energy. At the same time the total energy is not conserved and can become larger or smaller, depending on the problem.

\subsection{Conservation of Flux.}
\label{sec:appA}

The Floquet modes, $f_p$, fully describe the scattering process. Since they always contain scattering states, there should be no probability to find a particle close to the potential at $t\to\infty$ (assuming a square-integrable wave function at $t\to-\infty$): the total outgoing flux should be equal to the total incoming flux. Note that since the states with $p^2-\omega(m+n)<0$ do not give any contribution to the fluxes, the particle will leave these modes after some time. Physically it is easily understood, since a particle in these modes can undergo a transition to a scattering state and leave the range of the potential. This appendix 
shows the conservation of flux explicitly in a time-independent picture. Let us start with the scattering off zero-range potential described by Eq.~(\ref{eq:set_zero}), for which the flux is 
\begin{equation}
\vec j = i (\psi \vec \nabla \psi^*-\psi^* \vec \nabla \psi).
\end{equation}
The incoming flux is along the $x$ axis and amounts to $2p$.
The outgoing flux consists of the two parts: the first is along the $x$ direction and equals to $\sum_{p_n\geq0} 2p_n|C_n|^2$. The second piece is along the $(-x)$
 direction and amounts to $\sum_{p_n\geq0} 2 p_n |C_n-\delta_{n,0}|^2$. Let us show that from Eq.~(\ref{eq:set_zero}) it follows that 
\begin{equation}
\sum_{p_n\geq 0}p_n|C_n|^2=p \mathrm{Re} C_0,
\label{eq:fluxzerorange}
\end{equation}
which means that the total flux is conserved. To this end, we multiply
Eq.~(\ref{eq:set_zero}) with~$C_n^{*}$,
\begin{equation}
2 p_n |C_n|^2=2p\delta_{n,0} C_n^*-ig_0 |C_n|^2 - \frac{i g_1}{2}(C_{n+1}+C_{n-1})C_n^*.
\end{equation}
Next we conjugate Eq.~(\ref{eq:set_zero}) and then multiply
with~$C_n$,
\begin{equation}
2 p_n^* |C_n|^2=2p\delta_{n,0} C_n+ig_0 |C_n|^2 + \frac{i g_1}{2}(C^*_{n+1}+C^*_{n-1})C_n.
\end{equation}
Now we add these two equations and sum over all states
\begin{equation}
\sum (p_n+p_n^*)|C_n|^2=2p \mathrm{Re} C_0.
\end{equation}
Since $p_n+p_n^*$ is non-zero only for the scattering states we obtain Eq. (\ref{eq:fluxzerorange}). 
Similar steps can be taken to show that the total flux is conserved for any short-range potential. 

\bibliographystyle{unsrt}
\bibliography{bib}

\end{document}